\documentclass[aps,tightenlines,nofootinbib,preprint,
superscriptaddress,groupedaddress,epsfig]{revtex4}

\usepackage{graphicx}

\begin{document}

\title{Decay constants of $P$-wave mesons} \vspace{4mm}

\author{Guo-Li Wang}
\email{gl_wang@hit.edu.cn} \affiliation{Department of Physics,
Harbin Institute of Technology, Harbin, 150001, China}

\baselineskip=20pt

\begin{abstract}

Decay constants of $P$-wave mesons are computed in the framework
of instantaneous Bethe-Salpeter method (Salpeter method). By
analyzing the parity and possible charge conjugation parity, we
give the relativistic configurations of wave functions with
definite parity and possible charge conjugation parity. With these
wave functions as input, the full Salpeter equations for different
$P$-wave states are solved, and the mass spectra as well as the
numerical values of wave functions are obtained. Finally we
compute the leptonic decay constants of heavy-heavy and
heavy-light $^3P_0$, $^3P_1$ and $^1P_1$ states.
\end{abstract}


\maketitle

\section{Introduction}

The quantities of meson decay constants are important, since they
play important roles in many aspects, and the studies of them have
become hot topics in recent years
\cite{Adler,Pagels,Reinders,Allton,Roberts,narison,penin,yamada,ryan,bernard,juttner,wangzg}.
But most of these investigations are focused on the estimating
decay constants for $S$-wave mesons, and we lack the knowledge of
decay constants for $P$-wave mesons, there are only a few of
papers available \cite{cheng,veseli,yaouanc,hsieh,hwang,du}. We
present a careful study of the decay constants for heavy $^3P_0$,
$^3P_1$ and $^1P_1$ states including the relativistic corrections.

In previous letters \cite{previous}, the decay constants of
heavy-heavy and heavy-light pseudoscalar ($^1S_0$) and vector
($^3S_1$) mesons are calculated in the framework of relativistic
instantaneous Bethe-Salpeter method \cite{BS} (also called
Salpeter method \cite{salp}), good agreement of our predictions
with recent lattice, QCD sum rule, other relativistic model
calculations as well as available experimental data is found.

In this letter, we extend our previous analysis to include
$P$-wave mesons, present the calculations of decay constants for
heavy $P$-wave states in the framework of full Salpeter equation
which is a relativistic method. Based on the $S-L$ coupling
scheme, we analyze the parity and possible charge conjugation of
$^3P_0$, $^3P_1$ and $^1P_1$ bound states, and give general
formula for the wave functions which are in relativistic form with
definite parity and charge conjugation symmetry ($0^{++}$,
$1^{++}$, $1^{+-}$ for equal mass $^3P_0$, $^3P_1$ and $^1P_1$
bound states, and $0^+$, $1^+$, $1^+$ for un-equal mass bound
states). Then with these wave functions as input, we solve the
full Salpeter equations, obtain the mass spectra and numerical
values of wave functions for different $P$-wave states. Finally,
we compute the leptonic decay constants for heavy $^3P_0$, $^3P_1$
and $^1P_1$ states.

This letter is organized as following, in section II, we introduce
the relativistic Bethe-Salpeter equation and Salpeter equation. In
section III, we give the formula of relativistic wave functions
and decay constants for $P$-wave states. We solve the full
Salpeter equations, obtain the mass spectra and wave functions of
$P$-wave mesons. Finally, we use these relativistic wave functions
to calculate the decay constants of heavy $P$-wave mesons and show
the numerical results and discussion in section IV.

\section{Instantaneous Bethe-Salpeter Equation}

In this section, we briefly review the instantaneous
Bethe-Salpeter equation and introduce our notations, interested
reader can find the details in Ref. \cite{cskimwang}.

The Bethe-Salpeter (BS) equation is read as \cite{BS}:
\begin{equation}
(\not\!{p_{1}}-m_{1})\chi(q)(\not\!{p_{2}}+m_{2})=
i\int\frac{d^{4}k}{(2\pi)^{4}}V(P,k,q)\chi(k)\;, \label{eq1}
\end{equation}
where $\chi(q)$ is the BS wave function, $V(P,k,q)$ is the
interaction kernel between the quark and antiquark, and $p_{1},
p_{2}$ are the momenta of the quark 1 and anti-quark 2. The total
momentum $P$ and the relative momentum $q$ are defined as:
$$p_{1}={\alpha}_{1}P+q, \;\; {\alpha}_{1}=\frac{m_{1}}{m_{1}+m_{2}}~,$$
$$p_{2}={\alpha}_{2}P-q, \;\; {\alpha}_{2}=\frac{m_{2}}{m_{1}+m_{2}}~.$$

We divide the relative momentum $q$ into two parts,
$q_{\parallel}$ and $q_{\perp}$,
$$q^{\mu}=q^{\mu}_{\parallel}+q^{\mu}_{\perp}\;,$$
$$q^{\mu}_{\parallel}\equiv (P\cdot q/M^{2})P^{\mu}\;,\;\;\;
q^{\mu}_{\perp}\equiv q^{\mu}-q^{\mu}_{\parallel}\;.$$
Correspondingly, we have two Lorentz invariant variables:
\begin{center}
$q_{p}=\frac{(P\cdot q)}{M}\;, \;\;\;\;\;
q_{_T}=\sqrt{q^{2}_{p}-q^{2}}=\sqrt{-q^{2}_{\perp}}\;.$
\end{center}
When $\stackrel{\rightarrow}{P}=0$, they turn to the usual
component $q_{0}$ and $|\vec q|$, respectively.

After instantaneous approach, the kernel $V(P,k,q)$ takes the
simple form:
$$V(P,k,q) \Rightarrow V(k_{\perp},q_{\perp})\;.$$

Let us introduce the notations $\varphi_{p}(q^{\mu}_{\perp})$ and
$\eta(q^{\mu}_{\perp})$ for three dimensional wave function as
follows:
$$
\varphi_{p}(q^{\mu}_{\perp})\equiv i\int
\frac{dq_{p}}{2\pi}\chi(q^{\mu}_{\parallel},q^{\mu}_{\perp})\;,
$$
\begin{equation}
\eta(q^{\mu}_{\perp})\equiv\int\frac{dk_{\perp}}{(2\pi)^{3}}
V(k_{\perp},q_{\perp})\varphi_{p}(k^{\mu}_{\perp})\;. \label{eq5}
\end{equation}
Then the BS equation can be rewritten as:
\begin{equation}
\chi(q_{\parallel},q_{\perp})=S_{1}(p_{1})\eta(q_{\perp})S_{2}(p_{2})\;.
\label{eq6}
\end{equation}
The propagators of the two constituents can be decomposed as:
\begin{equation}
S_{i}(p_{i})=\frac{\Lambda^{+}_{ip}(q_{\perp})}{J(i)q_{p}
+\alpha_{i}M-\omega_{i}+i\epsilon}+
\frac{\Lambda^{-}_{ip}(q_{\perp})}{J(i)q_{p}+\alpha_{i}M+\omega_{i}-i\epsilon}\;,
\label{eq7}
\end{equation}
with
\begin{equation}
\omega_{i}=\sqrt{m_{i}^{2}+q^{2}_{_T}}\;,\;\;\;
\Lambda^{\pm}_{ip}(q_{\perp})= \frac{1}{2\omega_{ip}}\left[
\frac{\not\!{P}}{M}\omega_{i}\pm
J(i)(m_{i}+{\not\!q}_{\perp})\right]\;, \label{eq8}
\end{equation}
where $i=1, 2$ for quark and anti-quark, respectively,
 and
$J(i)=(-1)^{i+1}$. Here $\Lambda^{\pm}_{ip}(q_{\perp})$ satisfy
the relations:
\begin{equation}
\Lambda^{+}_{ip}(q_{\perp})+\Lambda^{-}_{ip}(q_{\perp})=\frac{\not\!{P}}{M}~,\;\;
\Lambda^{\pm}_{ip}(q_{\perp})\frac{\not\!{P}}{M}
\Lambda^{\pm}_{ip}(q_{\perp})=\Lambda^{\pm}_{ip}(q_{\perp})~,\;\;
\Lambda^{\pm}_{ip}(q_{\perp})\frac{\not\!{P}}{M}
\Lambda^{\mp}_{ip}(q_{\perp})=0~. \label{eq9}
\end{equation}

Introducing the notations $\varphi^{\pm\pm}_{p}(q_{\perp})$ as:
\begin{equation}
\varphi^{\pm\pm}_{p}(q_{\perp})\equiv
\Lambda^{\pm}_{1p}(q_{\perp})
\frac{\not\!{P}}{M}\varphi_{p}(q_{\perp}) \frac{\not\!{P}}{M}
\Lambda^{{\pm}}_{2p}(q_{\perp})\;, \label{eq10}
\end{equation}
and we have
$$
\varphi_{p}(q_{\perp})=\varphi^{++}_{p}(q_{\perp})+
\varphi^{+-}_{p}(q_{\perp})+\varphi^{-+}_{p}(q_{\perp})
+\varphi^{--}_{p}(q_{\perp})
$$
With contour integration over $q_{p}$ on both sides of
Eq.(\ref{eq6}), we obtain:
$$
\varphi_{p}(q_{\perp})=\frac{
\Lambda^{+}_{1p}(q_{\perp})\eta_{p}(q_{\perp})\Lambda^{+}_{2p}(q_{\perp})}
{(M-\omega_{1}-\omega_{2})}- \frac{
\Lambda^{-}_{1p}(q_{\perp})\eta_{p}(q_{\perp})\Lambda^{-}_{2p}(q_{\perp})}
{(M+\omega_{1}+\omega_{2})}\;,
$$
and the full Salpeter equation:
$$
(M-\omega_{1}-\omega_{2})\varphi^{++}_{p}(q_{\perp})=
\Lambda^{+}_{1p}(q_{\perp})\eta_{p}(q_{\perp})\Lambda^{+}_{2p}(q_{\perp})\;,
$$
$$(M+\omega_{1}+\omega_{2})\varphi^{--}_{p}(q_{\perp})=-
\Lambda^{-}_{1p}(q_{\perp})\eta_{p}(q_{\perp})\Lambda^{-}_{2p}(q_{\perp})\;,$$
\begin{equation}
\varphi^{+-}_{p}(q_{\perp})=\varphi^{-+}_{p}(q_{\perp})=0\;.
\label{eq11}
\end{equation}

The normalization condition for BS wave function is:
\begin{equation}
\int\frac{q_{_T}^2dq_{_T}}{2{\pi}^2}Tr\left[\overline\varphi^{++}
\frac{{/}\!\!\!
{P}}{M}\varphi^{++}\frac{{/}\!\!\!{P}}{M}-\overline\varphi^{--}
\frac{{/}\!\!\! {P}}{M}\varphi^{--}\frac{{/}\!\!\!
{P}}{M}\right]=2P_{0}\;. \label{eq12}
\end{equation}

 In our model, Cornell
potential, a linear scalar interaction plus a vector interaction
is chosen as the instantaneous interaction kernel $V$
\cite{cskimwang}:
$$V(\vec q)=V_s(\vec q)
+\gamma_{_0}\otimes\gamma^0 V_v(\vec q)~,$$
\begin{equation}
V_s(\vec q)=-(\frac{\lambda}{\alpha}+V_0) \delta^3(\vec
q)+\frac{\lambda}{\pi^2} \frac{1}{{(\vec q}^2+{\alpha}^2)^2}~,
~~V_v(\vec q)=-\frac{2}{3{\pi}^2}\frac{\alpha_s( \vec q)}{{(\vec
q}^2+{\alpha}^2)}~,\label{eq16}
\end{equation}
where the coupling constant $\alpha_s(\vec q)$ is running:
$$\alpha_s(\vec q)=\frac{12\pi}{27}\frac{1}
{\log (a+\frac{{\vec q}^2}{\Lambda^{2}_{QCD}})}~,$$ and the
constants $\lambda$, $\alpha$, $a$, $V_0$ and $\Lambda_{QCD}$ are
the parameters that characterize the potential.

\section{Relativistic Wave Functions and Decay Constants}
In this section, by analyzing the parity and possible charge
conjugation parity of corresponding bound state, we give a formula
for the wave function that is in a relativistic form with definite
parity and possible charge conjugation parity symmetry.
\subsection{ Wave function for $^3P_0$ state}

The general form for the relativistic Salpeter wave function of
$^3P_0$ state, which $J^P=0^+$ (or $J^{PC}=0^{++}$ for equal mass
system), can be written as:
\begin{equation}\varphi_{0^{+}}(q_{\perp})=
f_1(q_{\perp}){\not\!q}+f_2(q_{\perp})\frac{{\not\!P}
{\not\!q}_{\perp}}{M} +f_3(q_{\perp})M+f_4(q_{\perp}){\not\!P}.
\end{equation}
 The equations
\begin{equation}
\varphi^{+-}_{0^{+}}(q_{\perp})=\varphi^{-+}_{0^{+}}(q_{\perp})=0\;
\end{equation}
give the constraints on the components of the wave function, so we
have the relations
$$f_3(q_{\perp})=\frac{f_1(q_{\perp})q_{\perp}^2(m_1+m_2)}
{M(\omega_1\omega_2+m_1m_2+q_{\perp}^2)},~~~
f_4(q_{\perp})=\frac{f_2(q_{\perp})q_{\perp}^2(\omega_1-\omega_2)}
{M(m_1\omega_2+m_2\omega_1)}.$$ Then there are only two
independent wave functions $f_1(q_{\perp})$ and $f_2(q_{\perp})$
been left, from Eq. (\ref{eq11}), we obtain two coupled integral
equations, by solving them we obtain the numerical results of mass
spectra and wave functions, interesting reader can find the
details of this method in Ref. \cite{cskimwang} or Ref.
\cite{changwang}.

In our calculation, we choose the center-of-mass system of the
corresponding state, so $q_{\parallel}$ and $q_{\perp}$ turn to
the usual components $(q_0, {\vec 0})$ and $(0,{\vec q})$,
$\omega_{1}=(m_{1}^{2}+{\vec q}^{2})^{1/2}$ and
$\omega_{2}=(m_{2}^{2}+{\vec q}^{2})^{1/2}$. The normalization
condition for the $^3P_0$ wave function is:
 \begin{equation}
\int \frac{d{\vec q}}{(2\pi)^3}\frac{16f_1f_2
\omega_1\omega_2{\vec q}^2}{m_1\omega_2+m_2\omega_1}=2M.
 \end{equation}

\subsection{Decay constant of $^3P_0$ state}

The decay constant $F_{^3P_0}$ of scalar $^3P_0$ meson is defined
as
\begin{eqnarray}
\langle0|\bar{q_1}\gamma_\mu(1-\gamma_5) q_{2} |{^3P_0}\rangle
&\equiv& F_{^3P_0}P_\mu,
\end{eqnarray}
which can be written in the language of the Salpeter wave
functions as:

$$\langle0|\bar{q_1}\gamma_\mu(1-\gamma_5) q_{2} |{^3P_0}\rangle =
\sqrt{N_c}\int Tr \left[ \varphi_{0^+}({\vec
q})\gamma_{\mu}(1-\gamma_5)\right] \frac{d\vec q}{(2\pi)^3} $$
$$
=\sqrt{N_c}\int\frac{d{\vec q}}{(2\pi)^3} Tr \left[\left(
f_1{\not\!q}+f_2\frac{{\not\!P}{\not\!q}_{\perp}}{M}
-\frac{f_1{\vec q}^2(m_1+m_2)} {\omega_1\omega_2+m_1m_2-{\vec
q}^2} -\frac{f_2{\vec q}^2(\omega_1-\omega_2)\not\!P}
{M(m_1\omega_2+m_2\omega_1)}\right)\gamma_\mu\right]$$
\begin{eqnarray}
=4\sqrt{N_c}\int\frac{d{\vec q}}{(2\pi)^3}\left( -\frac{f_2{\vec
q}^2(\omega_1-\omega_2)} {M(m_1\omega_2+m_2\omega_1)}\right)P_\mu
\end{eqnarray}

Therefore, we have
\begin{eqnarray}
F_{^3P_0} =\frac{4\sqrt{N_c}}{M}\int\frac{d{\vec q}}{(2\pi)^3}
\frac{f_2{\vec q}^2(\omega_2-\omega_1)}
{(m_1\omega_2+m_2\omega_1)}\end{eqnarray}

\subsection{Wave function for $^3P_1$ state}

The general form for the Salpeter wave function of $^3P_1$ state,
which $J^P=1^+$ (or $J^{PC}=1^{++}$ for equal mass system), can be
written as:
\begin{equation}\varphi_{1^{+}}(q_{\perp})=i\varepsilon_{\mu\nu\alpha\beta}
P^{\nu}q_{\perp}^{\alpha}\epsilon^{\beta}\left[f_1M\gamma^{\mu}+
f_2{\not\!P}\gamma^{\mu}+f_3{\not\!q}_{\perp}\gamma^{\mu}
+if_4\varepsilon^{\mu\rho\sigma\delta}
q_{\perp\rho}P_{\sigma}\gamma_{\delta}\gamma_{5}/M \right]/M^2.
\end{equation}
 The equations
\begin{equation}
\varphi^{+-}_{1^{+}}(q_{\perp})=\varphi^{-+}_{1^{+}}(q_{\perp})=0\;
\end{equation}
give the constraints on the components of the wave function
$$f_3(q_{\perp})=\frac{f_1(q_{\perp})M(m_1\omega_2-m_2\omega_1)}
{q_{\perp}^2(\omega_1+\omega_2)},~~~
f_4(q_{\perp})=\frac{f_2(q_{\perp})M(-\omega_1\omega_2+m_1m_2+q_{\perp}^2)}
{q_{\perp}^2(m_1+m_2)}$$

The normalization condition for the $^3P_1$ wave function is:
 \begin{equation}
\int \frac{d{\vec q}}{(2\pi)^3}\frac{32f_1f_2
\omega_1\omega_2(\omega_1\omega_2-m_1m_2+{\vec
q}^{2})}{3(m_1+m_2)(\omega_1+\omega_2)}=2M.
 \end{equation}

\subsection{Decay constant of $^3P_1$ state}

The decay constant $F_{^3P_1}$ is defined as
\begin{eqnarray}
\langle0|\bar{q_1}\gamma_\mu(1-\gamma_5) q_{2}
|^3P_1,\epsilon\rangle &\equiv&
F_{^3P_1}M{\epsilon}^{\lambda}_\mu,
\end{eqnarray}
and can be formulated using the Salpeter wave function as:
\begin{eqnarray}\langle0|\bar{q_1}\gamma_\mu(1-\gamma_5) q_{2}
|^3P_1,\epsilon\rangle = \sqrt{N_c}\int Tr \left[
\varphi_{1^+}({\vec q})\gamma_{\mu}(1-\gamma_5)\right] \frac{d\vec
q}{(2\pi)^3},
\end{eqnarray}
then we have
\begin{eqnarray}
F_{^3P_1} =\frac{8\sqrt{N_c}}{3M}\int\frac{d{\vec q}}{(2\pi)^3}
\frac{f_2(\omega_1\omega_2-m_1m_2+{\vec
q}^{2})}{(m_1+m_2)}.\end{eqnarray}

\subsection{Wave function for $^1P_1$ state}

The general form for the Salpeter wave function of $^1P_1$ state,
which $J^P=1^+$ (or $J^{PC}=1^{+-}$ for equal mass system), can be
written as:
\begin{equation}\varphi_{1^{+}}(q_{\perp})=
q_{\perp}\cdot{\epsilon}^{\lambda}_{\perp}\left[
f_1(q_{\perp})+f_2(q_{\perp})\frac{{\not\!P}}{M}
+f_3(q_{\perp})\frac{{\not\!q_{\perp}}}{M}+
f_4(q_{\perp})\frac{\not\!{P}{\not\!q}}{M^2}\right]\gamma_5.
\end{equation}
 The equations
\begin{equation}
\varphi^{+-}_{1^{+}}(q_{\perp})=\varphi^{-+}_{1^{+}}(q_{\perp})=0\;
\end{equation}
give the constraints on the components of the wave function
$$f_3(q_{\perp})=-\frac{f_1(q_{\perp})M(m_1-m_2)}
{(\omega_1\omega_2+m_1m_2-q_{\perp}^2)},~~~
f_4(q_{\perp})=-\frac{f_2(q_{\perp})M(\omega_1+\omega_2)}
{(m_1\omega_2+m_2\omega_1)}$$

The normalization condition for the $^1P_1$ wave function is:
 \begin{equation}
\int \frac{d{\vec q}}{(2\pi)^3}\frac{16f_1f_2
\omega_1\omega_2{\vec q}^2}{3(m_1\omega_2+m_2\omega_1)}=2M.
 \end{equation}

\subsection{Decay constant of $^1P_1$ state}

The decay constant $F_{^1P_1}$ is defined as
\begin{eqnarray}
\langle0|\bar{q_1}\gamma_\mu(1-\gamma_5) q_{2}
|^1P_1,\epsilon\rangle &\equiv&
F_{^1P_1}M{\epsilon}^{\lambda}_\mu,
\end{eqnarray}
which can be formulated as:
\begin{eqnarray}
\langle0|\bar{q_1}\gamma_\mu(1-\gamma_5) q_{2}
|^1P_1,\epsilon\rangle = \sqrt{N_c}\int Tr \left[
\varphi_{1^+}({\vec q})\gamma_{\mu}(1-\gamma_5)\right] \frac{d\vec
q}{(2\pi)^3},
\end{eqnarray}
finally, we obtain
\begin{eqnarray}
F_{^1P_1} =\frac{4\sqrt{N_c}}{3M}\int\frac{d{\vec q}}{(2\pi)^3}
\frac{f_1(m_1-m_2){\vec q}^{2}}{(\omega_1\omega_2+m_1m_2+{\vec
q}^{2})}.
\end{eqnarray}

\section{Numerical Results and Discussion}
In our method, there are some input parameters appearing in the
potential, we need to fix them when solving the full Salpeter
equations. Usually, we fixed the parameters by fitting the
experimental mass spectra for mesons, but for $P$-wave states, we
lack experimental data, so we adopt almost the same parameters as
in the $0^-$ states Ref.~\cite{cskimwang}, and only vary the
parameter $V_0$ by fitting the ground $P$-wave $c\bar c$ states,
$\chi_{c0}$, $\chi_{c1}$ and $h_c$. In previous letter
\cite{previous}, we found if we choose same parameters set, the
mass predictions of our model can not agree very well with
experimental data for pseudoscalar and vector mesons, we find the
same thing happens to the different $P$-wave states, so we vary
the only possible different parameter $V_0$ to fit the data.
 For $^3P_0$ states, we
choose $V_0=-0.566$ GeV, for $^3P_1$ states, $V_0=-0.452$ GeV, and
for $^1P_1$ states, $V_0=-0.437$ GeV. The values of other
parameters are same as in the $0^-$ states Ref.~\cite{cskimwang}:
$$
a=e=2.7183, \alpha=0.06 {\rm GeV}, \lambda=0.20 {\rm GeV}^2,
\Lambda_{QCD}=0.26 {\rm GeV}~ {\rm and}
$$
\begin{equation} m_b=5.224 {\rm GeV}, m_c=1.7553 {\rm GeV}, m_s=0.487 {\rm GeV},
m_d=0.311 {\rm GeV}, m_u=0.305 {\rm GeV}. \label
{para}\end{equation}

\begingroup
\begin{table*}[hbt]
\setlength{\tabcolsep}{0.5cm} \caption{\small Mass spectra in unit
of MeV for $c\bar c$ and $b\bar b$ P wave states.} \label{tab2}
\begin{tabular*}{\textwidth}{@{}c@{\extracolsep{\fill}}ccccc}
 \hline \hline
{\phantom{\Large{l}}}\raisebox{+.2cm}{\phantom{\Large{j}}}
&Ex$(c\bar c)$&$c\bar c$&Ex$(b\bar b)$&$b\bar b$\\
\hline

{\phantom{\Large{l}}}\raisebox{+.2cm}{\phantom{\Large{j}}}
$1~^3P_0$&3415.2&3415.9&9859.9&9860.1    \\

{\phantom{\Large{l}}}\raisebox{+.2cm}{\phantom{\Large{j}}}
$2~^3P_0$&&3831.1&10232.1&10223.9    \\

{\phantom{\Large{l}}}\raisebox{+.2cm}{\phantom{\Large{j}}}
$3~^3P_0$&&4132.4&&10497.0    \\

{\phantom{\Large{l}}}\raisebox{+.2cm}{\phantom{\Large{j}}}
$4~^3P_0$&&4369.4&&10719.2    \\\hline

{\phantom{\Large{l}}}\raisebox{+.2cm}{\phantom{\Large{j}}}
$1~^3P_1$&3510.6&3510.9&9892.7&9892.1    \\

{\phantom{\Large{l}}}\raisebox{+.2cm}{\phantom{\Large{j}}}
$2~^3P_1$&&3923.1&10255.2&10255.0    \\

{\phantom{\Large{l}}}\raisebox{+.2cm}{\phantom{\Large{j}}}
$3~^3P_1$&&4222.0&&10527.4    \\

{\phantom{\Large{l}}}\raisebox{+.2cm}{\phantom{\Large{j}}}
$4~^3P_1$&&4456.9&&10750.0    \\\hline

{\phantom{\Large{l}}}\raisebox{+.2cm}{\phantom{\Large{j}}}
$1~^1P_1$&3524.4&3524.4&&9900.4    \\

{\phantom{\Large{l}}}\raisebox{+.2cm}{\phantom{\Large{j}}}
$2~^1P_1$&&3935.8&&10262.6    \\

{\phantom{\Large{l}}}\raisebox{+.2cm}{\phantom{\Large{j}}}
$3~^1P_1$&&4234.2&&10534.6    \\

{\phantom{\Large{l}}}\raisebox{+.2cm}{\phantom{\Large{j}}}
$4~^1P_1$&&4468.7&&10757.1    \\

\hline\hline
\end{tabular*}
\end{table*}
\endgroup

We show our theoretical predictions of mass spectra for $c\bar c$
states up to the $4P$ states as well as the experimental data in
Table I. One can see that, our predictions for mass splitting are,
$2P-1P\simeq 410$ MeV, $3P-2P\simeq 300$ MeV, $4P-3P\simeq 235$
MeV; the predicted mass for state $2^3P_1$ is $3923$ MeV, this is
a little smaller but consist with the traditional prediction of
potential model , which is about $50$ MeV higher than the one of
new state $X(3872)$. We show the predicted mass spectra for other
states In Table II, the interesting quantity is also the mass
splitting between the first radial excited state and ground state,
for all the $^3P_0$, $^3P_1$ and $^1P_1$ states, $2P-1P\simeq 330$
MeV for $u\bar b$, $2P-1P\simeq 345$ MeV for $u\bar c$,
$2P-1P\simeq 362$ MeV for $s\bar b$ and $2P-1P\simeq 381$ MeV for
$s\bar c$.

\begingroup
\begin{table*}[hbt]
\setlength{\tabcolsep}{0.5cm} \caption{\small Mass spectra in unit
of MeV for heavy P wave states.} \label{tab1}
\begin{tabular*}{\textwidth}{@{}c@{\extracolsep{\fill}}cccccccc}
 \hline \hline
{\phantom{\Large{l}}}\raisebox{+.2cm}{\phantom{\Large{j}}}
&$1~^3P_0$&$2~^3P_0$&$1~^3P_1$&$2~^3P_1$&$1~^1P_1$&$2~^1P_1$\\
\hline {\phantom{\Large{l}}}\raisebox{+.2cm}{\phantom{\Large{j}}}
$c\bar b$&6728.7&7127.8&6829.5&7225.3&6845.1&7239.6    \\

{\phantom{\Large{l}}}\raisebox{+.2cm}{\phantom{\Large{j}}}
$s\bar b$&5767.2&6130.3&5830.9&6192.2&5836.4&6197.3     \\

{\phantom{\Large{l}}}\raisebox{+.2cm}{\phantom{\Large{j}}}
$d\bar b$&5667.9&5998.9&5711.7&6042.8&5709.2&6041.4     \\

{\phantom{\Large{l}}}\raisebox{+.2cm}{\phantom{\Large{j}}}
$u\bar b$&5664.8&5994.1&5707.6&6037.2&5704.7&6035.5     \\

{\phantom{\Large{l}}}\raisebox{+.2cm}{\phantom{\Large{j}}}
$s\bar c$&2386.5&2767.4&2447.8&2827.3&2449.8&2830.4     \\

{\phantom{\Large{l}}}\raisebox{+.2cm}{\phantom{\Large{j}}}
$d\bar c$&2273.1&2619.2&2314.1&2661.3&2307.6&2657.7     \\

{\phantom{\Large{l}}}\raisebox{+.2cm}{\phantom{\Large{j}}}
$u\bar c$&2269.3&2613.7&2309.5&2655.0&2302.5&2651.1     \\

\hline\hline
\end{tabular*}
\end{table*}
\endgroup

We also calculate the mass spectra for $P$-wave $b \bar b$ system,
as argued in Ref. \cite{twophoton}, there are double heavy $b$
quarks, and the flavor $N_f=4$, so we have to choose a new set of
parameters as well as smaller value of coupling constant. We
change the previous scale parameters to $\Lambda_{QCD}=0.20$ GeV,
$m_b=5.13$ GeV which have been adopted in Ref. \cite{twophoton},
 choose $V_0=-0.553$ GeV for $^3P_0$ states, $V_0=-0.521$ GeV for $^3P_1$
states, $V_0=-0.514$ GeV for $^1P_1$ states, and other parameters
are not changed. With this set of parameters, the coupling
constant at the scale of bottom quark mass is
$\alpha_s(m_b)=0.23$. The numerical results and experimental data
of mass spectra for $b\bar b$ system are also shown in Table I.
One can see that our predictions, $2P-1P\simeq 363$ MeV, can fit
the experimental data very well, and our mass splitting
prediction, $3P-2P\simeq 273$ MeV, $4P-3P\simeq 223$ MeV.

\begingroup
\begin{table*}[hbt]
\setlength{\tabcolsep}{0.5cm} \caption{\small Decay constants in
unit of MeV for P wave $c\bar c$ and $b\bar b$ states.} \label{}
\begin{tabular*}{\textwidth}{@{}c@{\extracolsep{\fill}}cccccccc}
 \hline \hline
{\phantom{\Large{l}}}\raisebox{+.2cm}{\phantom{\Large{j}}}
&$n~^3P_0$&$n~^1P_1$&$1~^3P_1$&$2~^3P_1$&$3~^3P_1$&$4~^3P_1$\\
\hline {\phantom{\Large{l}}}\raisebox{+.2cm}{\phantom{\Large{j}}}
$c\bar c$&0&0&206&$-$207&199&$-$189    \\

{\phantom{\Large{l}}}\raisebox{+.2cm}{\phantom{\Large{j}}}
$b\bar b$&0&0&129&$-$131&126&$-$121    \\

\hline\hline
\end{tabular*}
\end{table*}
\endgroup

\begingroup
\begin{table*}[hbt]
\setlength{\tabcolsep}{0.5cm} \caption{\small Decay constants in
unit of MeV for heavy P wave states.} \label{}
\begin{tabular*}{\textwidth}{@{}c@{\extracolsep{\fill}}cccccccc}
 \hline \hline
{\phantom{\Large{l}}}\raisebox{+.2cm}{\phantom{\Large{j}}}
&$1~^3P_0$&$2~^3P_0$&$1~^3P_1$&$2~^3P_1$&$1~^1P_1$&$2~^1P_1$\\
\hline {\phantom{\Large{l}}}\raisebox{+.2cm}{\phantom{\Large{j}}}
$c\bar b~~$~~&~~88&$-$85&160&$-$165&50&$-$49    \\

{\phantom{\Large{l}}}\raisebox{+.2cm}{\phantom{\Large{j}}}
$s\bar b~~$~~&~~140&$-$130&157&$-$156&76&$-$71     \\

{\phantom{\Large{l}}}\raisebox{+.2cm}{\phantom{\Large{j}}}
$d\bar b~~$~~&~~145&$-$129&150&$-$144&76&$-$70     \\

{\phantom{\Large{l}}}\raisebox{+.2cm}{\phantom{\Large{j}}}
$u\bar b~~$~~&~~145&$-$128&150&$-$143&76&$-$70     \\

{\phantom{\Large{l}}}\raisebox{+.2cm}{\phantom{\Large{j}}}
$s\bar c~~$~~&~~112&$-$91&219&$-$204&62&$-$50     \\

{\phantom{\Large{l}}}\raisebox{+.2cm}{\phantom{\Large{j}}}
$d\bar c~~$~~&~~132&$-$102&212&$-$190&72&$-$56     \\

{\phantom{\Large{l}}}\raisebox{+.2cm}{\phantom{\Large{j}}}
$u\bar c~~$~~&~~133&$-$102&211&$-$189&72&$-$56     \\

\hline\hline
\end{tabular*}
\end{table*}
\endgroup

Becides the mass spectra, we also obtained the relativistic wave
functions for heavy mesons when solving the full Salpeter
equation. With these wave functions, we calculated the decay
constants for heavy-heavy and heavy-light $P$ wave mesons. In
Table III, we show our estimates of decay constants for $c \bar c$
and $b \bar b$ $^3P_1$-wave systems up to third radial excited
states, for $^3P_0$ and $^1P_1$ equal-mass states, the decay
constants vanish. The relative sign between the $F_{1P}$ and
$F_{2P}$ are minus, this come from the behavior of wave function,
since there is a node in the ${2P}$ radial wave function, the
minus sign means the dominant contribution come from the part
after the node. Our predictions show that, the decrease of the
numerical value of decay constant for higher excited state is not
evident comparing with the lower excited state, for example,
$F_{3P}-F_{1P}=7$ MeV for $^3P_1$ $c\bar c$ system, and
$F_{3P}-F_{1P}=3$ MeV for $^3P_1$ $b\bar b$ system.

\begingroup
\squeezetable
\begin{table*}[hbt]
\setlength{\tabcolsep}{0.5cm} \caption{\small Decay constants in
unit of MeV for heavy P wave states in different models.} \label{}
\begin{tabular*}{\textwidth}{@{}c@{\extracolsep{\fill}}ccccccccccc}
 \hline \hline
{\phantom{\Large{l}}}\raisebox{+.2cm}{\phantom{\Large{j}}}
&&$1~^3P_0$&&&$1~^3P_1$&&&$1~^1P_1$&&\\
\hline {\phantom{\Large{l}}}\raisebox{+.2cm}{\phantom{\Large{j}}}
~~&~~ours&\cite{cheng}&\cite{veseli}&ours&\cite{cheng}&\cite{veseli}&ours&\cite{cheng}&\cite{veseli}
\\\hline

{\phantom{\Large{l}}}\raisebox{+.2cm}{\phantom{\Large{j}}}
$s\bar b~~$~~&~~140&&146&157&&181&76&&84     \\

{\phantom{\Large{l}}}\raisebox{+.2cm}{\phantom{\Large{j}}}
$u\bar b~~$~~&~~145&112&162&150&123&187&76&68&93     \\

{\phantom{\Large{l}}}\raisebox{+.2cm}{\phantom{\Large{j}}}
$s\bar c~~$~~&~~112&71&110&219&121&240&62&38&63     \\

{\phantom{\Large{l}}}\raisebox{+.2cm}{\phantom{\Large{j}}}
$u\bar c~~$~~&~~133&86&139&211&127&249&72&45&82     \\

\hline\hline
\end{tabular*}
\end{table*}
\endgroup

In Table IV, we show our estimates of decay constants for
unequal-mass $P$-wave ground and first radial excited states. It
is observed that the decay constants of $^3P_1$ states are much
larger than those of the corresponding $^1P_1$ states, while the
corresponding values for $^3P_0$ is between them.

For comparison, we show our predictions for decay constants and
other theoretical predictions \cite{cheng,veseli} in Table V. We
have changed results in Ref. \cite{veseli} from the $j-j$ coupling
scheme to $S-L$ coupling scheme by using the following equations
\cite{cheng,isgur}
\begin{equation}
|^1P_1\rangle=\sqrt{\frac{2}{3}}\,|P^{3/2}_1\rangle -{1\over
\sqrt{3}}\,|P^{1/2}_1\rangle,\qquad |^3P_1\rangle={1\over
\sqrt{3}}\,|P^{3/2}_1\rangle
+\sqrt{\frac{2}{3}}\,|P^{1/2}_1\rangle.
\end{equation}
Rough agreement can be found between the values of decay constants
estimated by different methods, this means we need more effort for
the knowledge of $P$-wave decay constants.

In conclusion, we estimated the decay constants for heavy $P$-wave
$^3P_0$, $^3P_1$ and $^1P_1$
mesons in the framework of the relativistic Bether-Salpeter method. \\

\noindent This work was supported in part by the National Natural
Science Foundation of China (NSFC) under Grant No. 10675038.
\\

\end{document}